# Energy Internet Routing using Quantum Optimization Algorithms

**Alireza Alamgir Tehrani[1], Mehrdad Boroushaki[1], Abbas Rajabi[1]**

[1]Department of Energy Engineering, Sharif University of Technology, Tehran, Iran
Corresponding Author: Mehrdad Boroushaki (boroushaki@sharif.edu)

## Abstract
The Energy Internet (EI) is a new concept aimed at enhancing the integration of renewable energy sources with the energy grid. Energy-efficient path selection in EI is NP-hard. This research presents an innovative Quadratic Unconstrained Binary Optimization (QUBO) and Ising Hamiltonian formulation for energy routing. The validation and scalability of the proposed formulation were evaluated by applying quantum-inspired annealing and quantum gate optimization to two case studies, a 9-node and a 30-node EI network. A comparative analysis was presented between classical optimization using the Dijkstra algorithm, optimization-based methods, and QAOA using the Qiskit Sampler Primitive, NumpyEigenSolver, and quantum-inspired annealing using the Ocean exact Solver, D-Wave Tabu Sampler, and D-Wave Simulated Annealing. Simulation results based on the proposed formulation agree with the exact solution, while the runtime of classical approaches is less than that of quantum approaches. However, the Simulated Annealing sampler offers the shortest runtime among all quantum methods.

## 1. Introduction

The rapid growth of variable renewable energy in recent years poses notable challenges. Inefficiencies in connecting and managing traditional energy grids have resulted in notable unresolved challenges, including imbalances due to the unpredictability and generation fluctuations [1][2]. In addition, the advancement of distributed energy systems, the transition from centralized to decentralized grids, and the resulting transmission losses have complicated energy distribution and supply-demand balancing, revealing the necessity for effective management and control of renewable energy sources. The advent of the Energy Internet (EI) offers a new framework and innovative approach to solving these problems [3].

Fig. 1 shows different components of the EI. EI is a network that integrates electrical devices, including renewable and energy resources, loads, storage systems, control, and management systems, to enhance energy infrastructure, mitigate renewable energy integration challenges, and reduce environmental pollution, operating as a unified and integrated system with interconnected components and a structure similar to the Internet [3]. These issues highlight the importance of addressing a key NP-hard problem in the EI, energy routing [4]. An efficient transmission path is crucial for optimizing transmitted power, thereby enhancing the overall performance of the system.

As the core component of the EI, the energy router links electrical devices in the network, managing both the bidirectional flow of energy and data. The energy router's main task is energy routing; this encompasses the identification of the optimal path for energy transfer from sources to loads to reduce transmission energy loss [5].

The energy routing problem involves solving three main protocols: the subscriber matching protocol, which determines the best producer for each consumer; the energy-efficient routing protocol, which finds an optimal path with minimum energy loss during transmission between producer-consumer pairs; and the transmission scheduling protocol, which prevents congestion or overflow in the system [6].
Quantum computing has recently garnered attention as a highly promising approach with substantial potential to tackle NP-hard problems. Quantum computing relies on quantum mechanical principles, enabling parallel processing and exploration of solution spaces simultaneously through the principle of quantum superposition in a way that cannot occur in classical methods [7].
There are two main types of quantum computers: circuit-based computers and quantum annealer computers. Circuit-based quantum computers, like those used by IBM using the Qiskit platform, operate through sequences of quantum gates to solve complex problems such as cryptography, optimization, and machine learning. Several quantum algorithms have been developed for solving optimization problems on circuit-based quantum computers, including the Quantum Approximate Optimization Algorithm (QAOA) and the Variational Quantum Eigensolver (VQE) algorithms [8].

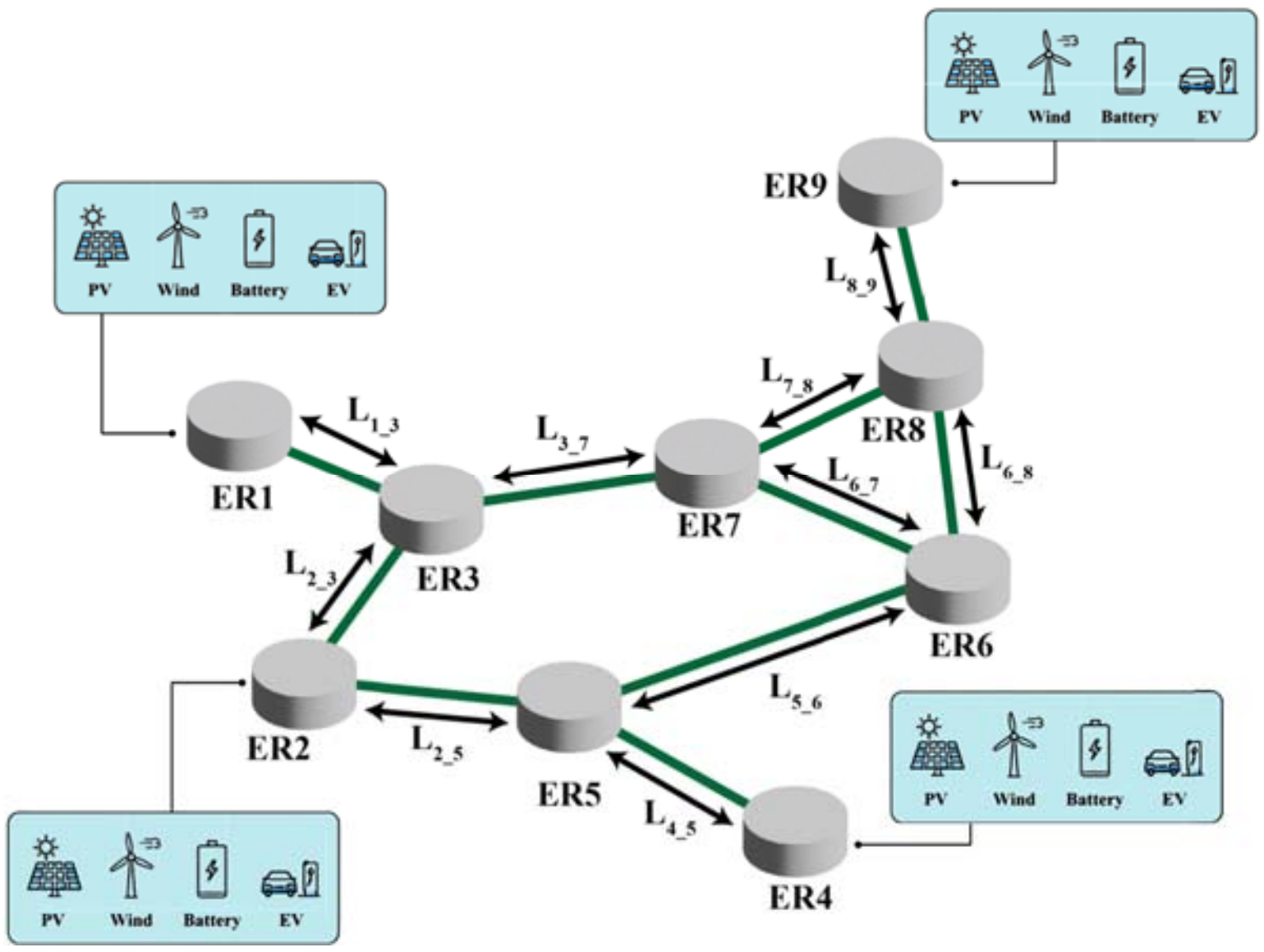

**Figure. 1**: Energy Internet components where each node is an energy router (ER)

The QAOA was the primary circuit-based quantum approach used in studies to solve the Vehicle Routing Problem (VRP) [9-13], while some works instead implemented the VQE, and all of these studies employed Quadratic Unconstrained Binary Optimization (QUBO) or Ising formulations for their models. A data-driven approach to QAOA was presented in [14] to identify maximum sections of power or data delivery in Distributed Energy Resource (DER)-dominant cyber-physical power systems.
Quantum annealer computers like the D-Wave, based on the Ocean platform, have been specially developed for solving optimization problems. Quantum annealing encodes the problem into a Hamiltonian and gradually evolves the system from an initial Hamiltonian to a problem Hamiltonian, allowing quantum tunneling to drive the system toward a low-energy ground state that represents the optimal or near-optimal solution [15].
A comparative analysis in [16] was performed on the quantum alternating operator ansatz and quantum annealing applied to 127-qubit random Ising models, considering both quadratic and cubic terms. The study used the IBM_Washington with 127 qubits and also D-Wave's Pegasus with 5,600 qubit quantum processors. The results showed that quantum annealing consistently yields near-optimal solutions and superior performance to QAOA.

A quantum annealing-based approach in [17] was proposed to optimize electricity surplus in transmission power networks, aiming to reduce costs and improve efficiency. The problem was formulated as a QUBO model for network partitioning and tested on both quantum and hybrid approaches. Analysis showed that hybrid quantum methodologies, implemented via D-Wave's hybrid constrained and binary quadratic model solvers, yielded superior results over classical methodologies, achieving reduced objective function values for a range of problem scales.
In [18], the article established the VRP as a maximum weight independent set problem and formulated it as a QUBO, and employed D-Wave's quantum annealer for a solution. Quantum annealing exhibited a shorter annealing time than the Gurobi Optimizer, but the runtime variability was greater across instances. The research in [19] described three methodologies to solve the single-source, single-destination shortest path in network problem, formulated as a QUBO problem. It revealed that a speed advantage may be presented compared to the classical algorithms in sparse or low-degree graphs.
A graph-theory-based model using Dijkstra's algorithm was proposed in [20] to solve the single-source single-load energy routing problem in the Energy Internet by calculating low-loss paths between energy source-load pairs. A graph theory-based routing algorithm was also developed in [21] to solve the power routing problem in residential multi-microgrid (MMG) systems. The method minimizes power losses by identifying the best low-loss, congestion-free, and high-reliability routes. An algorithm was introduced in [22], for energy routing in a semi-decentralized system. This algorithm minimizes energy losses during transmission through the identification of the best low-loss path across potential scenarios.
The three main protocols within the EI, subscriber matching, energy-efficient path selection, and transmission scheduling are addressed in [23]. The article employs meta-heuristic evolutionary algorithms, and the results exhibit promising quality in terms of energy losses, cost, and computation time for complex networks.
Further application of quantum computing in solving engineering problems is of great importance to demonstrate its promising applications. To the best of the authors' knowledge, no research work has been reported on the application of quantum optimization to the energy routing problem. This paper presents a novel application of quantum computing in solving engineering problems.
This manuscript focuses on solving the single-source, single-load energy routing problem, aiming to minimize routing losses using quantum optimization techniques. The main contributions of this manuscript are outlined as follows:

- Presenting a novel application of quantum computing in solving engineering problems.
- The energy routing problem in EI is formulated as a novel QUBO and Ising Hamiltonian formulation
- Evaluation of the single source-load energy-routing problem is performed on a 9-node EI network using classical, quantum-inspired annealing, and QAOA methods, and also extended to a 30-node EI network using classical and quantum-inspired annealing
- A comparative analysis of different quantum and classical approaches is presented, focusing on optimal paths, losses, and computational time.

## 2. Preliminaries

The following two sections outline the required preliminaries for quantum approaches to deepen understanding of the proposed formulation.

### 2.1 Quantum approximation optimization algorithm (QAOA)

QAOA is a variational hybrid quantum-classical algorithm designed to approximate solutions to optimization problems. This algorithm consists of two essential parts: the mixing Hamiltonian and the cost Hamiltonian. The QAOA ansatz is constructed by applying alternating layers of cost and mixing operators as follows [16]:

$$\left|\psi_p(\gamma,\beta)\right\rangle = U_M(\beta_P)U_c(\gamma_p)\ldots U_M(\beta_1)U_c(\gamma_1)|\psi_1\rangle \qquad (1)$$

where the depth of the ansatz circuit is *p,* and represents the number of alternating layers. The variational parameters $\gamma$ and $\beta$ control each layer through the implementation $U_c(\gamma_k) = e^{-i\gamma H_c}$ to apply the phases corresponding to the objective function of the energy routing problem and $U_M(\beta_k) = e^{-i\beta H_M}$ to reallocate amplitudes among basis states. The output is measured after $p$ layers, resulting in a classical bit string. The expected cost $\langle H_c\rangle = \left\langle\psi_p(\gamma,\beta)|H_c|\psi_p(\gamma,\beta)\right\rangle$ is minimized by optimizing variational parameters using classical solvers.
Fig. 4 illustrates the QAOA algorithm implemented using the Qiskit Sampler Primitive in the EI routing problem. The computational cost of classical variational parameter optimization rises considerably with increasing the number of layers $p$ [25, 26].

### 2.2 Quantum annealing

Quantum annealing is an approach used to solve optimization problems by exploiting quantum mechanical principles. The process utilizes quantum tunneling and adiabatic evolution to locate minimum energy solutions. Like QAOA, quantum annealing necessitates representing the problem within a QUBO or an Ising Hamiltonian model, denoted as the cost Hamiltonian, which functions as the objective function for quantum optimization [27, 28]. The evolution of the system in quantum annealing is determined by a time-dependent Hamiltonian, which is analogous to the QAOA cost and mixing operators.

$$H(t) = -A(t)\sum_{i=1}^{n}\sigma_i{}^x + B(t)H_c \qquad (2)$$

where the first term denotes the driver Hamiltonian, functioning similarly to the mixing operator in QAOA, and $H_c$ signifies the cost Hamiltonian associated with the Ising objective function.
The evolution begins with the dominance of the driver Hamiltonian, generating a uniform superposition of states, and progressively transitions towards the cost Hamiltonian, ensuring the system remains in or near the ground state of *H(t)* throughout the process, in accordance with the adiabatic theorem [18, 26]. Quantum annealing implements quantum tunneling to explore the solution space and helps the system escape from local minima more efficiently than classical thermal annealing.
At the end of the evolution, the qubits are measured in the computational basis to obtain the final solution. Quantum annealing often employs multiple annealing runs to sample a distribution of solutions, and the lowest-energy sample is selected as the candidate solution. The performance of a quantum annealing system relies on the length of annealing time $T$, the loading of the problem on the hardware graph, and control of environmental noise [29].

### 2.3 Energy routing formulation

Fig. 2 shows the 9-node EI network with node 1 as the source and node 9 as the load.

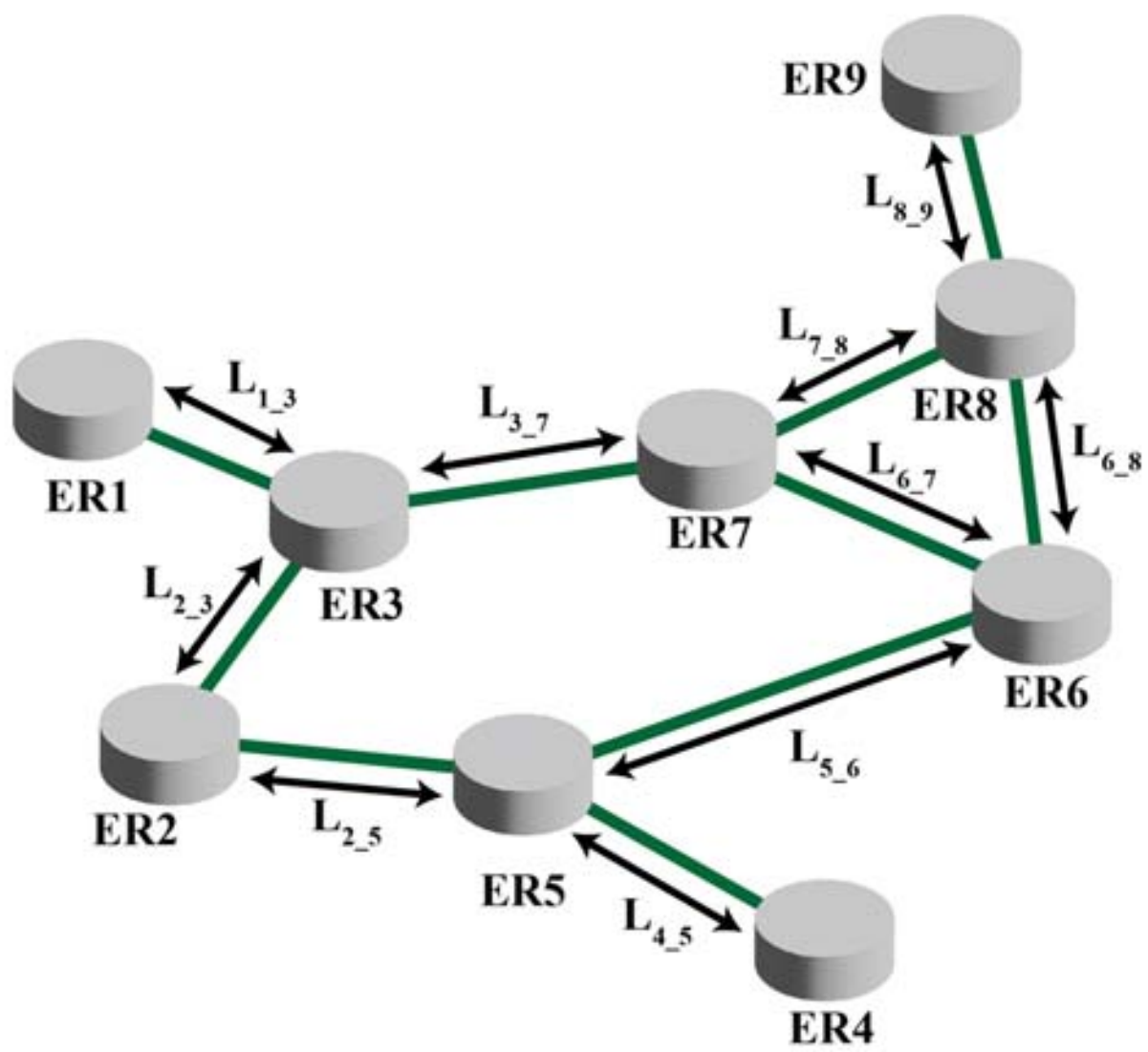

**Figure 2:** A sample 9-node EI network comprises interconnected Energy Routers, where ER1 serves as the energy source and ER9 as the load.

In the EI routing problem, the route with the minimum energy losses is used for energy transmission. The loss function $L_s$, for a given source-sink pair s, is defined as follows:

$$L_s = \sum_{K_s}^{N_s}(\sum W_i + \sum W_{ij}) \tag{3}$$

where $W_i$ is the power losses within the energy router *i* and is equal to.

$$W_i = (1 - \eta_i)P_{k_s} \tag{4}$$

where $\eta_i$ indicates the energy router's efficiency, $P_{k_s}$ denotes the power exchanged requested by the load-side energy router from the source energy router, specifically, the power to be dispatched through the *k-th* path. The conduction losses of the transmission line $W_{ij}$, is calculated by:

$$W_{ij} = \frac{R_{ij}}{V_{ij}}[(P_{k_s} + P_{ex})^2 - (P_{ex})^2] \tag{5}$$

where $R_{ij}$ symbolizes the resistance of the transmission line connecting the energy routers; $V_{ij}$ denotes the transmission line voltage between nodes *i* and *j*, and $P_{ex}$ refers to the power previously existing on the transmission line. Generally, the energy transmitted from all sources must be equivalent to the demand of the load $P_{d-s}$ as follows:

$$\sum_{k_s=1}^{N_s} P_{k_s} = P_{d-s} \tag{6}$$

## 3. Method

Fig. 3 shows the calculation path used for the EI routing problem, in which the problem is transformed into a QUBO model and then into an Ising optimization model. This model can be solved by gate-based quantum approaches via the QAOA method using the Qiskit Sampler Primitive, NumpyEigenSolver, and/or by quantum-inspired annealing approaches using the Ocean exact Solver, D-Wave Tabu Sampler, and D-Wave Simulated methods.

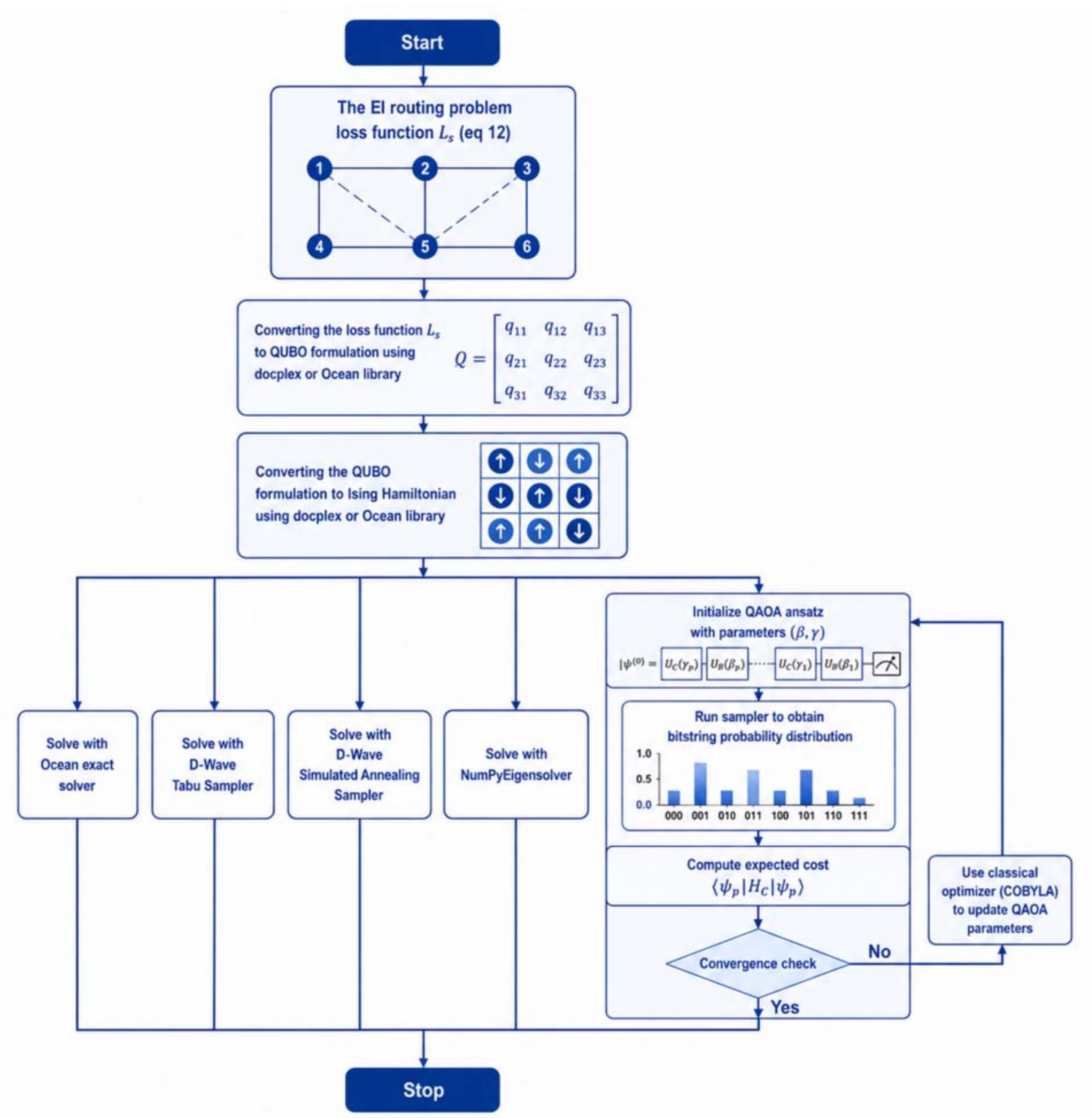

**Figure 3**: Flowchart of different quantum computation methods applied to the EI routing problem.

### 3.1 The QUBO formulation of the EI routing problem

To leverage quantum computing for optimization, the optimization problem must be transformed into QUBO form. The QUBO form can be used in both gate-based and annealing-based quantum computers. The energy routing problem is NP-hard. The objective function and constraints for this problem are reformulated using two binary variable sets for each node $x_i$ , and each edge $x_{ij}$.

$$L_s = \sum_{i \in N} w_i x_i + \sum_{(i,j) \in E} w_{ij} x_{ij} \quad (7)$$

The loss function $L_s$ represents the total losses along the paths, and $w_i, w_{ij}$ are energy losses in routers and transmission lines. The binary variables $x_i, x_{ij} \in \{0,1\}$ denote whether energy router $i$ or transmission line *(i,j)* is included in the optimal path, and are assigned to 1 if the component is part of the optimal path, and 0 otherwise. In the energy routing problem, the source node is restricted to a single output edge, the load node is restricted to a single input edge, and the rest of the nodes have an equal number of input and output edges as follows:

$$\sum_{j:(i,j) \in E} x_{ij} - \sum_{j:(j,i) \in E} x_{ji} = \begin{cases} 1 & if\ i = Source \\ -1 & if\ i = Load \\ 0 & otherwise \end{cases} \quad (8)$$

This ensures a single, directed path that begins at the source, concludes at the load, and preserves flow balance at all intervening nodes. Another constraint that must exist in the network is the consistency between the node and edge variables. Specifically, whenever an outgoing edge *(i, j)* or an incoming edge *(j, i)* is activated, both the corresponding nodes *i* and *j* must also be activated. These constraints guarantee that both nodes connected to a selected edge become activated. The mathematical formulation of these constraints applies to all nodes with $x_i \geq x_{ij}$ and $x_i \geq x_{ji}$ . However, to ensure consistency between node activation and edge activation, a node splitting technique can be used. In this technique, each node is replaced by two auxiliary nodes, known as the input node and the output node. All edges entering the node are directed to the input node, while all edges leaving the node are directed to the output node. The input and output nodes are then connected by an internal edge whose weight represents the amount of energy loss of the corresponding router. In this case, a node can be activated only by activating the internal edge connecting its input and output parts. In this way, the router loss is reflected in the objective function through the weight assigned to the internal edge. This reformulation results in a more compact model that can be useful when building a QUBO model of the network.

The QUBO problems are usually stated using the standard quadratic formulation as follows:

$$min(\, x^T Q x + c), x \in \{0,1\}^n \quad (9)$$

Given that $x = [x_1, x_2, \ldots, x_n]^T$ is a vector of binary variables, $Q \in R^{n \times n}$is an upper-triangular matrix of real-valued coefficients that encodes both quadratic and linear terms. The above QUBO formulation can be comprised into two components: a linear and a quadratic part.

$$H = \sum_{i=1}^{n} a_i x_i + \sum_{i=1}^{n} \sum_{j=1}^{n} b_{ij} x_i x_j \quad (10)$$

The nonlinearity of the EI routing problem makes its formulation in QUBO form challenging. To transform the problem into a QUBO formulation, one has to consider the power transmission $P_{k_s}$ in the network as a constant parameter, and express the non-binary variables as binary variables and reformulate the constraint optimization problem as an unconstrained problem.

Constraints are integrated into the objective function using the penalty-based approach. Thus, the objective function formulation is transformed as follows:

$$min\{c^T x: Ax = b, x \in \{0,1\}^n\} \rightarrow min\{c^T x + \rho \|Ax - b\|^2, x \in \{0,1\}^n\} \quad (11)$$

where the original constraint $Ax = b$ is incorporated into the objective function through a quadratic penalty term. Here, $x$ represents the decision variables, $c$ is the cost coefficient vector, A and b define the system constraints, and $\rho > 0$ is the penalty constant, which controls the degree of the violations.

The total loss $L_s$ by reformulating equations (7), (8) ,and the inequality constraints in the QUBO form comprises two parts: the objective function loss $L_{objective}$ and the constraints violation loss $L_{constraints}$, as follows:

$$L_s = L_{objective} + L_{constraints} \tag{12}$$

where the objective losses include energy router losses $w_i$ and transmission lines losses $w_{ij}$ as follows:

$$L_{objective} = \sum_{i \in nodes} w_i x_i + \sum_{(i,j) \in edges} w_{ij} x_{ij} \tag{13}$$

The constraints violation loss includes source, load, and path, as well as inequality constraints, as follows:

$$L_{constraints} = L_{source} + L_{load} + L_{path} + L_{inequality} \tag{14}$$

all of which are individually defined as follows:

$$L_{source} = \rho_1 \left(1 - \left(\sum x_{ij} - \sum x_{ji}\right)\right)^2 \tag{15}$$

$$L_{load} = \rho_2 \left(-1 - \left(\sum x_{ij} - \sum x_{ji}\right)\right)^2 \tag{16}$$

$$L_{path} = \rho_3 \left(\sum x_{ij} - \sum x_{ji}\right)^2 \tag{17}$$

$$L_{inequality} = \rho_4 \left(max(0, x_{ij} - x_i) + max(0, x_{ji} - x_i)\right) \tag{18}$$

This formulation represents the standard form of the QUBO problem. To facilitate the implementation of the optimization problem using quantum computing, one must transform this formulation into an Ising Hamiltonian form.

### 3.2 QUBO to Ising model expression of the energy routing

The Ising model is a classical mathematical model for calculating the total ferromagnetic energy via the following Ising Hamiltonian:

$$H = -\sum_{\langle i,j \rangle} J_{ij} \sigma_i \sigma_j + \sum_i h_i \sigma_i \tag{19}$$

where $\sigma_i \in \{-1, 1\}$, $J_{ij}$ symbolizes pairwise magnetic interactions, $h_i$ indicating the effect of an external magnetic field, $\sigma_i$ denoting a Pauli-Z gate applied to qubits, and $\sigma_i \sigma_j$ signifies the tensor product of two Pauli-Z gates equal to:

$$\sigma_i \sigma_j = \begin{bmatrix} 1 & 0 & 0 & 0 \\ 0 & -1 & 0 & 0 \\ 0 & 0 & -1 & 0 \\ 0 & 0 & 0 & 1 \end{bmatrix} \tag{20}$$

Despite its physics origins, this Ising Hamiltonian form is mathematically equivalent to a QUBO form. However, the key difference is the variable domains: QUBO uses binary variables $x_i \in \{0,1\}$ , while the Ising model uses Pauli variables $\sigma_i \in \{-1,1\}$ . The binary-variable transformation enables the mapping of the QUBO formulation to an Ising Hamiltonian form by

$$x_i = \frac{\sigma_i + 1}{2} \tag{21}$$

This transformation expresses the EI routing problem in the Ising Hamiltonian form, where the energy routing problem can be simulated physically using quantum systems by naturally evolving toward minimum energy configurations.

## 4. Results

The performance of single-source load energy routing is evaluated using simulation results from both classical and quantum-inspired optimization methods. Routing protocols used in EI are typically based on concepts of graph theory, which are outlined by a set$G(N, E, W)$. The notation uses $N$ for the nodes, $E$ for the edges or connections between nodes, and $W$ for the edge weights, which correspond to the distance or cost between two nodes. In the absence of a direct connection between two nodes, the weight $W$ is assigned to zero. Total transmission loss between a source and a load is categorized into two elements: energy losses due to power conversion in energy routers and conduction losses in transmission lines.

### 4.1 Quantum and classical implementation of the energy routing problem to a 9-node EI network

Fig. 4 presents a 9-node EI network graph structure. Each node $i$ in this graph is an energy router, and the edges represent the energy transmission lines. Node 1 is designated to be the source node, while node 9 is selected as the load node.

For the classical optimization approach, Dijkstra and the classical optimization-based methods were performed in Python 3.11.13 on Google Colab, using the glpk solver. For quantum optimization using gate-based quantum circuits, the QAOA algorithm was performed through the Qiskit Sampler Primitive and Qiskit's NumpyEigenSolvers methods. The quantum-inspired annealing optimization was performed through the Ocean exact solver, the D-Wave Tabu Sampler, and the D-Wave Simulated Annealing Sampler methods. The cost Hamiltonian for QAOA and quantum-inspired annealing will be detailed in future sections. The QAOA's circuit depth $p$ was set to 1, and the QAOA parameters were optimized using the COBYLA solver.

Table 1 shows the optimal routing path, associated energy losses, and computation time at a transmission power of 1 $kW$ for various classical and quantum approaches for the 9-node EI network, where all classical and quantum optimization methods successfully identified the optimal routing path (1–3–7–8–9), resulting in an identical energy loss of 74.687 $W$.

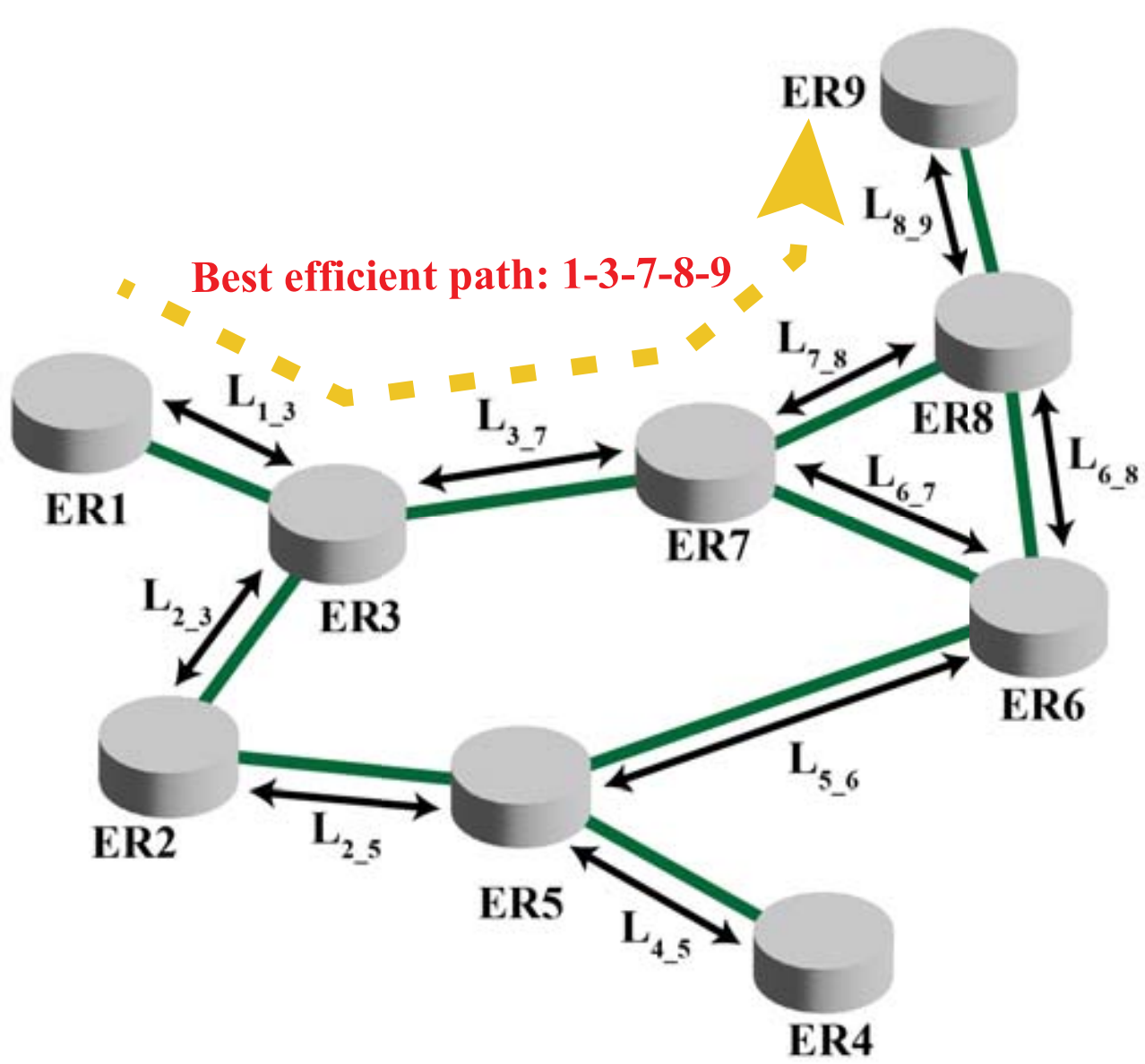

**Figure 4:** The optimal path in energy routing with node 1 as the source node and node 9 as the load node

This finding validates both the robustness of the methods and the correctness of the problem formulation. Dijkstra's algorithm, with a runtime of 25 $\mu s$, shows the optimal performance, considerably outperforming the other methods presented, owing to the effectiveness of classical deterministic graph-search algorithms for routing problems.

**TABLE 1.** Comparative results of various Energy Routing approaches for a 9-node EI network on a classical computer.

| Method | Optimal Path | Energy loss (W) | Computation Time (s) |
|---|---|---|---|
| Dijkstra Algorithm | 1-3-7-8-9 | 74.687 | **0.00002524** |
| Optimization-based method | 1-3-7-8-9 | 74.687 | 0.0377 |
| QAOA by Qiskit NumpyEigenSolver | 1-3-7-8-9 | 74.687 | 0.2095 |
| QAOA by Qiskit Sampler Primitive and the COBYLA classical solver | 1-3-7-8-9 | 74.687 | 539 |
| Quantum-inspired Annealing by Ocean exact solver | 1-3-7-8-9 | 74.687 | 209.7947 |
| Quantum-inspired Annealing by D-Wave Tabu Sampler | 1-3-7-8-9 | 74.687 | 2.1007 |
| Quantum-inspired Annealing by D-Wave Simulated Annealing Sampler | 1-3-7-8-9 | 74.687 | 0.02229 |

Similarly, the classical optimization-based method using the glpk mathematical solver provides the correct solution, too. However, the runtime is 0.0377 *s*, which is higher, stemming from the use of a complete programming model rather than a direct graph search.

All quantum simulations were executed on classical hardware, not on a real quantum computer. The reported times are increased due to classical processing and therefore do not represent the true performance on a real quantum processor. The QAOA implementation using the NumpyEigenSolver requires 0.2095 *s*.

The QAOA algorithm by Qiskit Sampler Primitive and the COBYLA solver method result in a runtime of 539 *s*. Results reveal the computational inefficiencies of the quantum gate optimization approach through the hybrid quantum-classical method by Qiskit Sampler Primitive against both classical methods. Both QAOA implementations successfully determined the accurate optimal path, while the classical NumpyEigenSolver achieved the optimal solution faster.

Among the quantum-inspired annealing methods, D-Wave simulated annealing Sampler required 0.02229 *s* as the lowest computation time. Therefore, while the simulated annealing Sampler offers a faster computation time than the QAOA simulation and the classical optimization-based method, it is still slower than Dijkstra's algorithm.

### 4.2 Quantum and classical implementation of the energy routing problem to a 30-node EI network

To examine the scalability of the proposed quantum formulation, one should apply it to a larger network. The study is further extended to a 30-node EI network with node 1 as the source node and node 30 as the load node [22]. Fig. 5 shows the optimal path by the yellow path calculated by different quantum and classical approaches.

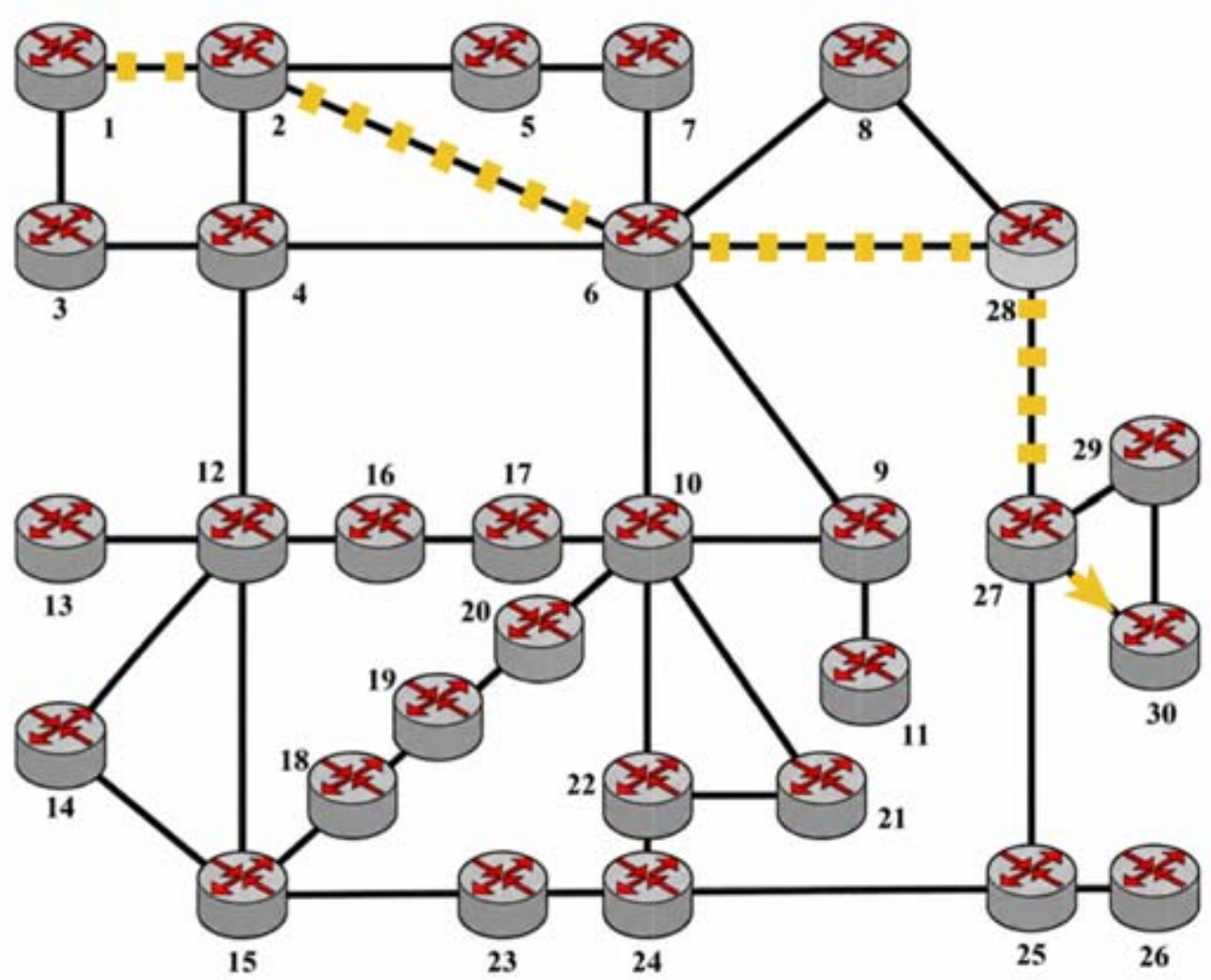

**Figure 5:** The optimal energy routing path between the source and the load in the 30-node EI network

Table 2 compares the results obtained for the 30-node EI network with different classical and quantum approaches. Comparative results show that the Dijkstra algorithm, with a runtime of 0.00005223 *s*, represents the most efficient method for addressing the energy routing problem, while the optimization-based method is the second most efficient method.

In contrast, the quantum-inspired annealing approach using D-Wave Simulated Annealing and Tabu Sampler methods exhibits the worst computational performance with average times of 1.391 and 2.1011 *s*, respectively.

Since the Quantum-inspired Annealing with Ocean exact solver method explores all possible paths, it requires exponential computation time. This method is infeasible and cannot be applied to a 30-node EI network.

Quantum modeling of this problem with 30 nodes and 41 edges required 112 qubits: 30 qubits for the nodes and 82 qubits for the bidirectional edges. This number of qubits exceeds the practical classical threshold for simulating arbitrary quantum circuits (approximately 50 qubits) because of the exponential scaling challenge [24]. Thus, the QAOA method could not be examined. All remaining methods identified the same optimal path between the designated source (node 1) and the load (node 30), namely 1-2-6-28-27-30, with a corresponding energy loss of 123.01 *W*.

**TABLE 2.** Comparative results of various Energy Routing approaches for a 30-node EI network on a classical computer.

| Method | Optimal Path | Energy loss (W) | Computation Time (s) |
|---|---|---|---|
| Dijkstra Algorithm | 1-2-6-28-27-30 | 123.01 | **0.00005223** |
| Optimization-based method | 1-2-6-28-27-30 | 123.01 | 0.05169 |
| QAOA by Qiskit NumpyEigenSolver | Infeasible | | |
| QAOA by Qiskit Sampler Primitive and the COBYLA classical solver | Infeasible | | |
| Quantum-inspired Annealing by Ocean exact solver | Infeasible | | |
| Quantum-inspired Annealing by D-Wave Tabu Sampler | 1-2-6-28-27-30 | 123.01 | 2.1011 |
| Quantum-inspired Annealing by D-Wave Simulated Annealing Sampler | 1-2-6-28-27-30 | 123.01 | 1.391 |

In conclusion, the application of quantum-inspired approaches solved on classical solvers represents efficiency for the EI routing problem, although classical Dijkstra algorithms maintain a significant legacy in best runtime performance.

All quantum methods in QAOA and quantum-inspired annealing successfully identified the optimal solution for small-scale problems, thus validating the proposed QUBO formulation and proving its correctness.

## 5. Discussion

This research presents a novel QUBO and Ising Hamiltonian formulation for the energy routing problem in EI. This is a crucial NP-hard, computationally demanding problem for large-scale energy systems. Converting the energy-efficient path selection for a single source–load pair to a QUBO formulation allows its direct implementation using the QAOA on gate-based quantum simulators in Qiskit and quantum-inspired annealing in Ocean platforms.

Comparative analysis of a 9-node with 10 edges Energy Internet topology reveals that classical methods, such as Dijkstra's algorithm and exact optimization, are highly efficient for small-to-medium-scale networks, identifying the optimal path (1–3–7–8–9) with energy losses of 74.687 $W$ at 1 $KW$ transmission power within a few milliseconds of runtime. Both quantum methods successfully identify the same optimal path and identical energy loss, thus validating the proposed QUBO and Ising formulation.

The performance of the two quantum approaches, quantum gate and quantum-inspired annealing, is clearly distinguishable. The results obtained from the QAOA algorithm using NumpyEigenSolver method have acceptable runtime. In contrast, the QAOA algorithm with the Qiskit Sampler primitive has the worst runtime.

In the quantum-inspired annealing–based optimization approach, three different solver/Sampler were implemented: the Ocean exact Solver, the D-Wave Tabu Sampler, and the D-Wave Simulated Annealing Sampler. Comparative analysis of the results shows that the Simulated Annealing Sampler outperformed the other methods in terms of runtime, while all methods found the same optimal path with the lowest energy loss.

To examine the scalability of the presented formulation, an EI routing problem with a larger dimension of 30 nodes, in addition to the problem with 9 nodes, was solved through classical approaches, including Dijkstra's, and the optimization-based methods. In the quantum approaches, the solution was limited to quantum-inspired annealing using D-Wave's Tabu and Simulated Annealing Sampler, while implementation of the QAOA algorithm was not feasible for the 30-node EI problem due to the large number of required qubits.

The process of transforming the problem into a QUBO format considerably increases the complexity. This can result in more computation and runtime. Given that the energy routing problem is inherently defined in the classical domain, it has been formulated in a quantum domain and subsequently solved using classical Samplers or solvers. This justifies why the classical methods outperform the quantum approaches.

The computational performance may be improved by running the problem on a physical Quantum Processing Unit (QPU), such as those from IBM or D-Wave. The fact that neither quantum-inspired annealing nor the QAOA method has been implemented on real quantum computers represents a significant shortcoming of the proposed formulation.

Quantum approaches require further development due to the nascent stage of quantum computing technologies. Despite the high expressibility of quantum gate-based approaches such as QAOA, classical approaches have already demonstrated practical effectiveness in addressing real-time energy routing problems.

## 6. Conclusions and Future Work

Quantum approaches, particularly in small networks, require further development due to the nascent stage of quantum computing technologies. Despite the high Expressibility of gate-based approaches such as QAOA, classical approaches have already demonstrated practical effectiveness in addressing real-time energy routing problems.

Future work will focus on expanding the proposed formulation to handle multi-source/multi-load scenarios, as well as integrating dynamic transmission scheduling and congestion control; the performance could be evaluated using real quantum annealing hardware rather than the exact solver. Due to the increasing scale and complexity of Energy Internet deployments, the inherent parallelism of quantum optimization techniques, particularly annealing-based methods, makes them highly suitable candidates for scalable, real-time energy routing in the Energy Internet of tomorrow.

## Author Contributions

A.A.T. carried out the main theoretical development, analytical calculations, numerical simulations, figure preparation, and manuscript writing. M.B. led and supervised the project, contributed to the theoretical framework and interpretation of the results, and edited and revised the manuscript. A.R. contributed to the theoretical analysis and interpretation of results. All authors discussed the results and approved the final manuscript.

## Data Availability

The dataset containing the energy router parameters and transmission line data used in this study is publicly available on Zenodo at: https://doi.org/10.5281/zenodo.20573872

## Code Availability

The codes developed for this project are available to reviewers via the following access link:

https://doi.org/10.5281/zenodo.20595672 on demand and will be made publicly available upon acceptance.